\documentclass{PoS}
\PoS{PoS(LAT2005)072}

\usepackage{amsmath,epsfig,amssymb}

\date{}

\def\lsi{\raise0.3ex\hbox{$<$\kern-0.75em\raise-1.1ex\hbox{$\sim$}}}
\def\gsi{\raise0.3ex\hbox{$>$\kern-0.75em\raise-1.1ex\hbox{$\sim$}}}

\newcommand{\bq}{\begin{eqnarray}}
\newcommand{\eq}{\end{eqnarray}}
\newcommand{\bqa}{\begin{eqnarray}}
\newcommand{\eqa}{\end{eqnarray}}

\newcommand{\be}{\begin{equation}}
\newcommand{\ee}{\end{equation}}
\newcommand{\bea}{\begin{eqnarray}}
\newcommand{\eea}{\end{eqnarray}}
\newcommand{\half}{\frac{1}{2}}

\title{Dynamical twisted mass fermions}

\ShortTitle{Dynamical twisted mass fermions}

\author{Federico Farchioni\footnote{Speaker}, Peter Hofmann, Gernot M\"unster\\
        Institut f\"ur Theoretische Physik, Universit\"at M\"unster,
        Wilhelm-Klemm-Str. 9, 48149 M\"unster, Germany\\
        E-mail: \email{farchion, munsteg@uni-muenster.de, hofmann@muenster.de}}

\author{Karl Jansen, Mauro Papinutto, Andrea Shindler, Urs Wenger$^*$, 
Ines Wetzorke$^*$\\\\
       NIC, Platanenallee 6, 15738 Zeuthen, Germany\\
E-mail: \email{karl.jansen, mauro.papinutto, andrea.shindler, urs.wenger, ines.wetzorke@desy.de}}

\author{Istvan Montvay, Enno E.~Scholz, Naoya Ukita$^*$\\
        DESY, Notkestr. 85, 22607 Hamburg, Germany\\
        E-mail: \email{istvan.montvay, enno.e.scholz, naoya.ukita@desy.de}}

\author{Luigi Scorzato\\
        Institut f\"ur Physik, Humboldt-Universit\"at zu Berlin,
        Newtonstr. 15, 12489 Berlin, Germany\\
        E-mail: \email{scorzato@physik.hu-berlin.de}}

\author{Carsten Urbach\\
        NIC, Zeuthen and Institut f\"ur Theoretische Physik, 
        Freie Universit\"at Berlin, Arnimallee 14, 14195 Berlin, Germany\\
        E-mail: \email{urbach@physik.fu-berlin.de}}

\abstract{
We summarize four contributions about dynamical twisted mass fermions.
The resulting report covers results for $N_f=2$ obtained from 
three different gauge actions, namely the
standard Wilson plaquette gauge action \cite{Wetzorke:proc}, the DBW2 
\cite{Ukita:proc} and the tree-level Symanzik improved~\cite{Wenger:proc}
gauge action. In addition, first results for $N_f=2+1+1$ flavours of twisted
mass fermions are discussed. 
}

\FullConference{XXIIIrd International Symposium on Lattice Field Theory\\
                 25-30 July 2005\\
                 Trinity College, Dublin, Ireland}

\begin{document}

\section{Introduction}

Within the framework of Symanzik's improvement programme \cite{Symanzik:1983dc,Symanzik:1983gh} 
the emphasis in lattice QCD has been on the construction and 
testing of fermion actions to avoid $O(a)$ lattice spacing artefacts. 
Since the lattice artefacts induced by the gauge action start with
$O(a^2)$, the need of also exploring different gauge actions has not been 
considered as equally urgent. However, although this point of view is certainly
valid in the asymptotic regime of small lattice spacing $a$, the situation 
is different when values of the lattice spacing, say, 
$a>0.1$~fm are considered. In this region higher order lattice artefacts 
my play an important role.

In the case of Wilson fermions the explicit breaking of chiral symmetry
leads~\cite{Sharpe:1998xm} to the appearance of a strong 
first order phase transition that severely affects the numerical 
simulations~\cite{Farchioni:2004us,Farchioni:2004ma}.  
As we will demonstrate in this contribution, the choice of the
gauge action can have a strong effect on the strength of this phase transition
and therefore it becomes important to consider both, the fermion and the 
gauge action to find a suitable lattice QCD action for simulations of 
dynamical quarks. 

The twisted mass fermions approach~\cite{Frezzotti:1999vv} provides, among other
advantages, the ideal framework for the investigation of the 
zero-temperature phase diagram of lattice QCD with Wilson fermions,
see refs.~\cite{Frezzotti:2002iv,Frezzotti:2004pc,Shindler:proc}
for reviews on twisted mass fermions in present and past conferences.

\begin{figure}[htb]
\vspace{-0.0cm}
\begin{center}
\epsfig{file=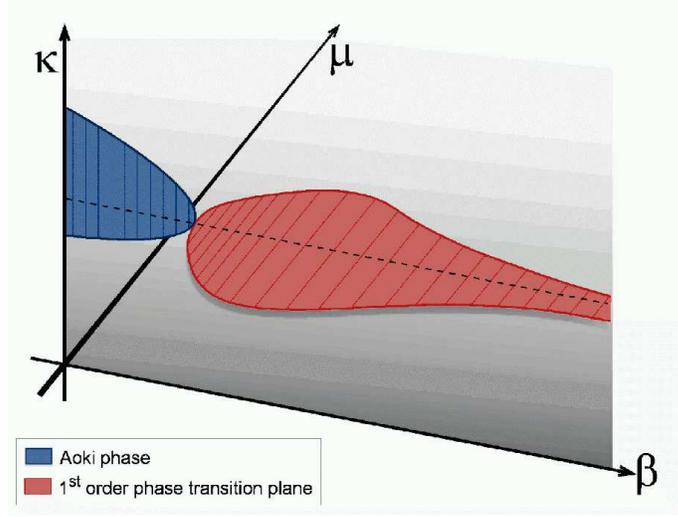,angle=0,width=0.6\linewidth}
\end{center}
\vspace{-0.5cm}
\caption{Current knowledge of the lattice QCD zero-temperature phase diagram with Wilson fermions
as a function of the inverse gauge coupling $\beta\propto 1/g^2$, the hopping parameter 
$\kappa$ and the twisted mass parameter $\mu$.
\label{fig:phase}}
\end{figure}

Our present understanding can be summarized as follows, see also
fig.~\ref{fig:phase}.
For values of the lattice spacing much coarser than $a=0.15$~fm 
an unphysical phase appears for small quark masses, the
Aoki phase~\cite{Aoki:1984qi,Ilgenfritz:2003gw,Sternbeck:2003gy}.
The transition from the standard lattice QCD phase to the Aoki phase
is of second order: here the charged pions become massless.
The Aoki phase only extends in the untwisted quark mass direction
being absent for non vanishing twisted quark mass $\mu$. 
For smaller values of the lattice spacing a first order phase transition
appears \cite{Farchioni:2004us,Farchioni:2004ma,Farchioni:2004fs,Farchioni:2005tu},
this time extending in the twisted mass direction $\mu$. The phase transition
occurs at vanishing (untwisted) quark mass and 
separates the phases with opposite signs of the quark mass. 
This first order phase transition is reminiscent of the continuum 
phase transition when the quark mass is changed from positive to negative values
with the corresponding jump of the scalar condensate as the order parameter of
spontaneous chiral symmetry breaking. 
The generic phase structure of lattice QCD 
was discussed in refs.~\cite{Farchioni:2004us,Farchioni:2004ma,Farchioni:2004fs}.

The appearance of the first order phase transition 
has serious consequences, since in such a scenario the pion mass
$m_\pi$ cannot be made arbitrarily small but assumes a minimal value,
$m_\pi^\mathrm{min}$, which may be about $500\ \mathrm{MeV}$ and hence
it becomes impossible to work close to the physical value of the pion
mass. 
It therefore becomes important to 
understand the phase structure of lattice QCD
as a pre-requisite before starting large scale
simulations. 
Our collaboration has performed a detailed study 
of the phase diagram of lattice QCD. We used Wilson fermions with and without
twisted mass parameter and studied the phase diagram as a function of the 
lattice spacing for the Wilson plaquette, the DBW2 and the tree-level 
Symanzik improved gauge actions. As we will show in this contribution, 
the strength of the first order phase transition, namely the
size of $m_\pi^\mathrm{min}$, depends strongly on the choice of 
the gauge action when comparable physical situations are tested. 

Since lattice chiral perturbation theory ($\chi$PT) 
\cite{Sharpe:1998xm,Munster:2004am,Scorzato:2004da,Sharpe:2004ny,Sharpe:2004ps,Aoki:2004ta}
predicts a weakening of the first order phase
transition toward the continuum limit, it is interesting to check
this prediction and, in particular, to investigate   
how fast the transition weakens when the continuum limit is
approached. The answer to the latter question will naturally depend on
the choice of the actions that are used for the gauge and the fermion
fields.
Moreover, the predictions of $\chi$PT for e.g.
the quark mass dependence of the pion mass and the pseudoscalar decay constant at 
such a phase transition can be directly confronted to results from numerical simulations.

This contribution is organized as follows. In section~2 we will give 
the definitions of the actions we will use. In sections~3,~4 and 5 we will 
give our results for the Wilson plaquette, the DBW2 and the tree-level 
Symanzik improved actions, respectively. In section~6 we will confront our lattice data
with lattice $\chi$PT. In section~7 we will provide
first results from simulations with $N_f=2+1+1$ flavors of quarks.
Finally, we will conclude in section~8.

\section{Lattice action}
\label{sec:action}

The lattice action for a doublet of degenerate twisted mass Wilson fermions 
(in the so-called ``twisted basis'') reads 
\be\label{eq:ferm_action}
S_q = \sum_x \left\{ 
\left( \overline{\chi}_x [\mu_\kappa + i\gamma_5\tau_3a\mu ]\chi_x \right)
- \half\sum_{\mu=\pm 1}^{\pm 4}
\left( \overline{\chi}_{x+\hat{\mu}}U_{x\mu}[r+\gamma_\mu]\chi_x \right)
\right\} \ ,
\ee
with $\mu_\kappa \equiv am_0 + 4r = 1/2\kappa$,
$r$ the Wilson-parameter, set in our simulations to $r=1$, $am_0$
the bare ``untwisted'' quark mass in lattice units ($\kappa$ is the conventional hopping parameter) 
and
$\mu$ the twisted quark mass; we also define $U_{x,-\mu} = U_{x-\hat{\mu},\mu}^\dagger$ and
$\gamma_{-\mu}=-\gamma_\mu$.

For the gauge sector we consider the one-parameter family of actions
including planar rectangular $(1\times 2)$ Wilson loops 
($U_{x\mu\nu}^{1\times 2}$):
\be\label{eq:gauge_action}
S_g = \beta\sum_{x}\left(c_{0}\sum_{\mu<\nu;\,\mu,\nu=1}^4
\left\{1-\frac{1}{3}\,{\rm Re\,} U_{x\mu\nu}^{1\times 1}\right\}
+c_{1}\sum_{\mu\ne\nu;\,\mu,\nu=1}^4
\left\{1-\frac{1}{3}\,{\rm Re\,} U_{x\mu\nu}^{1\times 2}\right\}
\right) \ ,
\ee
with the normalization condition $c_{0}=1-8c_{1}$.
We consider in this contribution the three cases: i.) Wilson plaquette gauge action, $c_{1}=0$,
ii.) DBW2 gauge action~\cite{Takaishi:1996xj}, $c_{1}=-1.4088$,
iii.) tree-level Symanzik improved gauge action (tlSym)~\cite{Weisz:1982zw}, $c_1 = -1/12$.

\section{Wilson plaquette gauge action}

The first action we investigate here is the Wilson plaquette gauge action.
We studied the lattice spacing dependence of several quantities 
\cite{Farchioni:2005tu} keeping the twisted mass and the 
lattice size roughly fixed
to $r_0 \mu \approx 0.03$ ($r_0$ being the Sommer parameter
\cite{Sommer:1993ce}, we assume $r_0=0.5$~fm throughout this contribution) 
and $L \approx 2$~fm. We varied $\beta=6/g^2$ in the range
$\beta=5.1-5.3$. For each value of the hopping parameter we performed 
a hot and a cold
start in order to check for co-existing values of physical observables
which we have chosen as the pion mass $m_{\pi}$ and the rho meson mass $m_\rho$,
extracted from the usual correlation functions. 
Another quantity is the amplitude $f^{PS}_\chi$, which reduces to the pion
decay constant $f_\pi$ for $\mu=0$, defined as
\bqa
f^{PS}_\chi = \frac{r_{AP}}{m_{\pi}}\langle0|P(0)|\pi\rangle
\equiv \frac{r_{AP}}{m_{\pi}} g_{\pi}\ ,
\mbox{~~~where~~~}
&r_{AP}& = \frac{\langle0|A_0(0)|\pi\rangle}{\langle0|P(0)|\pi\rangle}
\eqa
is extracted from the asymptotic behavior of the ratio of the axialvector-pseudoscalar 
correlator $C_{AP}(x_0)$ to the pseudoscalar-pseudoscalar correlator $C_{PP}(x_0)$ 
(cf. ref~\cite{Farchioni:2005tu} and references therein for details):
$
\frac{C_{AP}(x_0)}{C_{PP}(x_0)} = r_{AP} \tanh\left[m_{\pi}(T/2-x_0)\right]
\; . 
$
Moreover we measure the untwisted PCAC quark mass $m^{PCAC}_\chi$:
\bqa
m_\chi^{PCAC} = \frac{f^{PS}_\chi\;m_{\pi}^2}{2 \langle0|P(0)|\pi\rangle}
&\equiv& \frac{f^{PS}_\chi\;m_{\pi}^2}{2 g_{\pi}}\;.
\label{eqpcacmass}
\eqa
Note that the physical decay constant (PCAC quark mass) is obtained
by a combination of $f^{PS}_\chi$ and the twist angle $\omega$ ($m^{PCAC}_\chi$
and the twisted mass $\mu$), see also section~4.
\begin{figure}[htb]
\vspace{-0.0cm}
\begin{center}
\epsfig{file=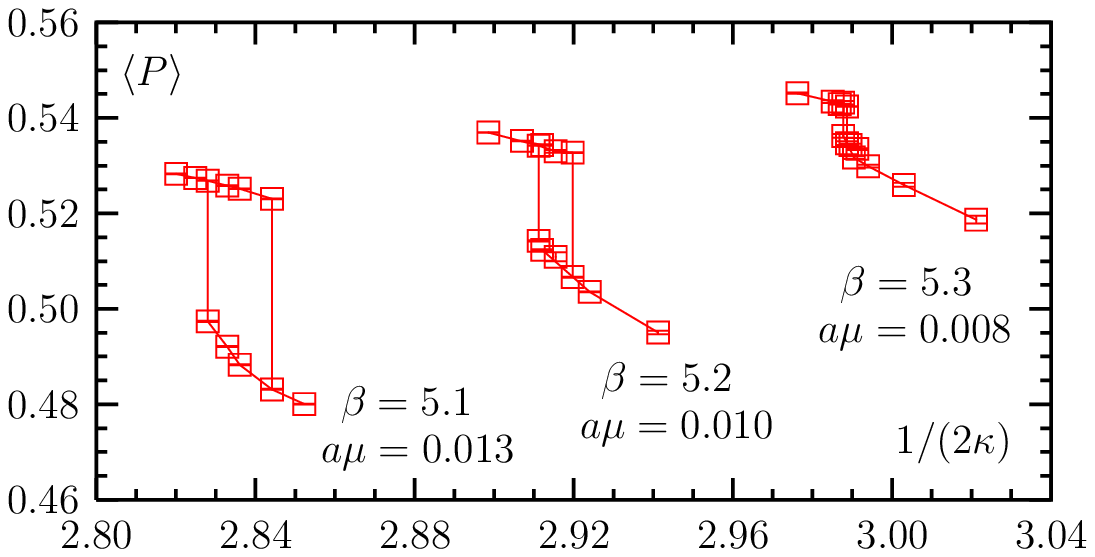,width=0.45\linewidth}
\epsfig{file=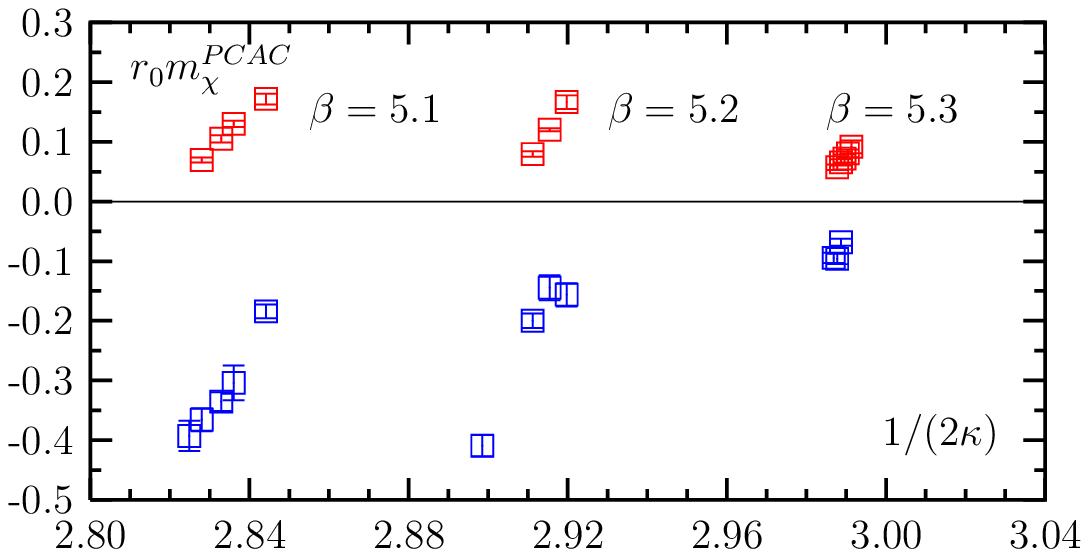,width=0.45\linewidth}
\end{center}
\vspace{-0.7cm}
\caption{Lattice spacing dependence of the plaquette expectation value
~(left) and the PCAC quark mass~(right).
\label{fig:plaq}}
\end{figure}

\begin{figure}[htb]
\vspace{-0.0cm}
\begin{center}
\epsfig{file=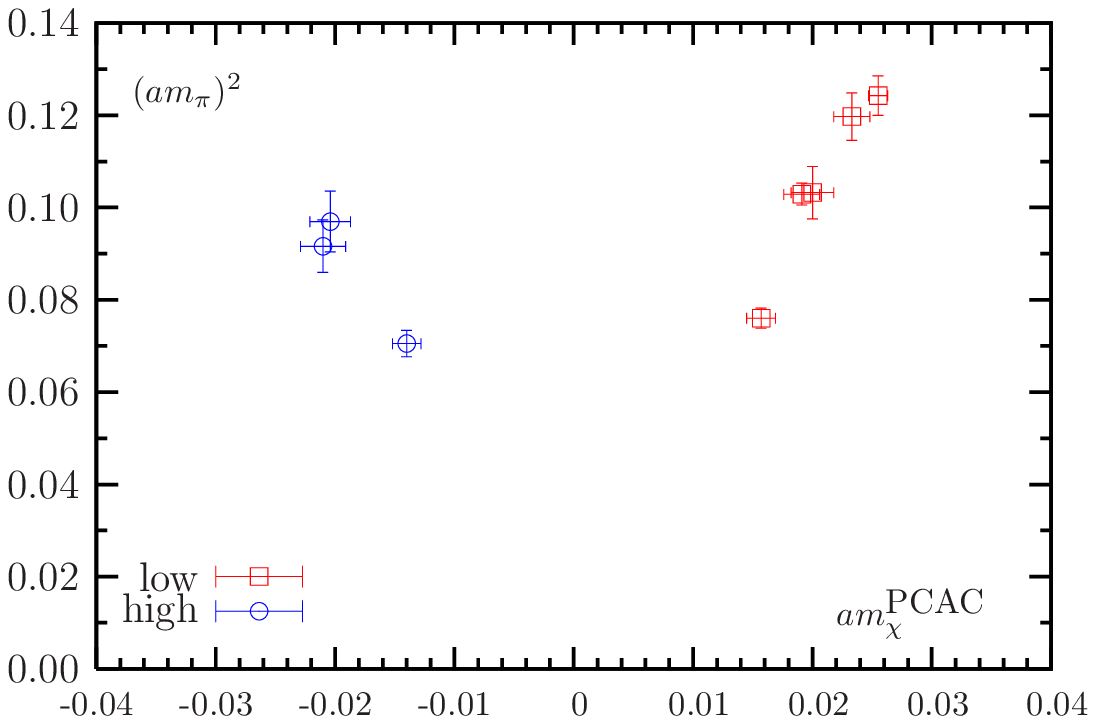,width=0.45\linewidth}
\epsfig{file=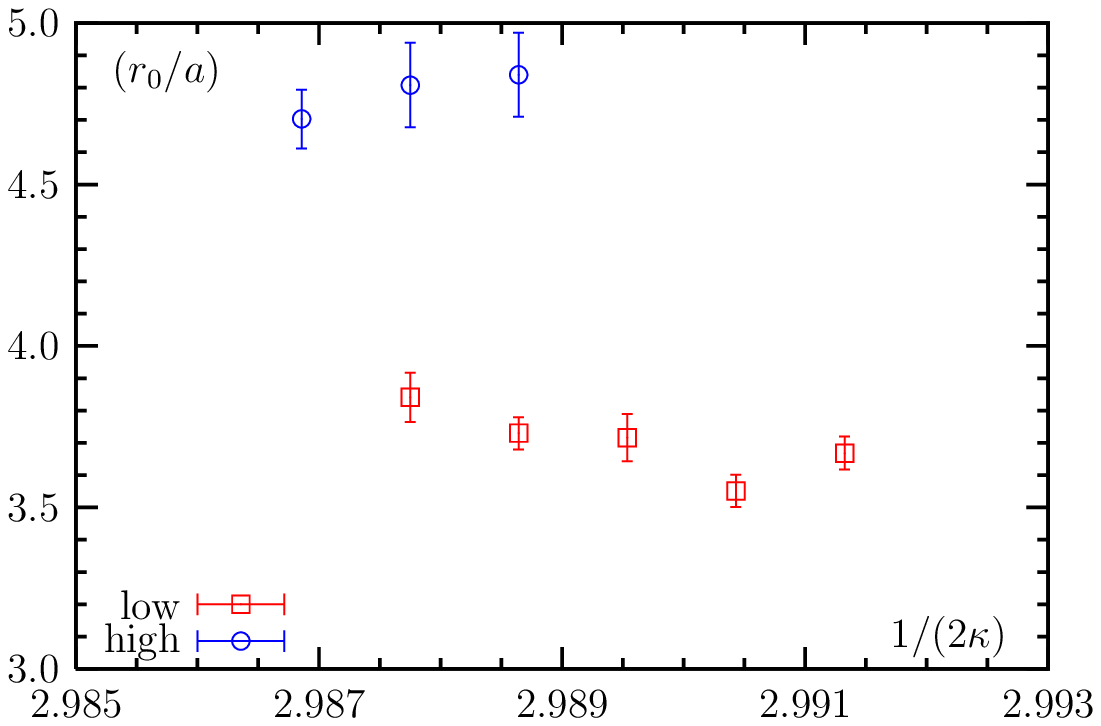,width=0.45\linewidth}
\end{center}
\vspace{-0.7cm}
\caption{Pion mass vs. PCAC quark mass~(left) and mass dependence of 
$r_0/a$~(right) at $\beta=5.3$. ``Low'' and ``high'' refer
to the low plaquette (positive quark mass) and the high plaquette (negative quark mass)
phase, respectively. 
\label{fig:pion}}
\end{figure}

The plaquette expectation value 
shows the typical behavior at a first order phase transition
with meta-stable branches, illustrated in fig.~\ref{fig:plaq}~(left). For the
PCAC quark mass the two branches correspond to positive and negative quark
mass, respectively (see fig.~\ref{fig:plaq}~(right)). 
In both cases the gap decreases
from $\beta=5.1$ to 5.3 and the meta-stability region in $1/2\kappa$ shrinks.
For the pion mass shown in fig.~\ref{fig:pion}~(left) this fact has the
important consequence that only a minimal pion mass $\sim500$ MeV can be
reached. If one attempts to lower the mass further by tuning $\mu$ or $\kappa$,
a jump to the other phase occurs. 
In addition, the scale parameter $r_0/a$ is quite different in the two
phases and its mass dependence is non-negligible, as
illustrated in fig.~\ref{fig:pion}~(right).

In the light of these observations the interesting question arises
at which $\beta$-value or lattice spacing we can safely simulate 
pions with masses as light as  $\sim300$ MeV in order to make contact with 
$\chi$PT. With the present data it is very difficult to
make an extrapolation, since there is a large ambiguity in the
determination of the minimal pion mass, the lever arm is very short
due to the small range of $\beta$=5.1-5.3 covered and furthermore
due to the large difference of the scale parameter $r_0/a$ in the two
phases. Nevertheless, a qualitative estimate would be a lattice spacing
of 0.07-0.1~fm, where one should be able to reach small pion masses
without being affected by the first order phase transition.
This makes even $L=2$~fm simulations, e.g.~for a detailed scaling study,
very demanding. Therefore alternative gauge actions that lead to a
reduced strength of the first order phase transition are investigated in
the next sections.

\subsection{Scaling behavior}

Here we study the scaling behavior with our current data for the
Wilson plaquette action, also including the data from the DBW2 gauge action, see 
sec.~\ref{sec:omega_ren}.
To this end we express the physical quantities in dimensionless variables.  
We first define a  reference quark mass by $(r_0 m_{\pi})^2=1.5$.
At this reference quark mass, we then determine\footnote{We consider only 
the positive quark mass phase which corresponds
to standard lattice QCD.}
 reference values for 
the quantities we are interested in, i.e. $r_0/a$, $m_\chi^{PCAC}$, 
$m_{\pi}$, $m_\rho$ and $f^{PS}_\chi$.
This allows to define dimensionless ratios such as for the quark mass,
$\sigma=\frac{m_\chi^{PCAC}}{m_\chi^{PCAC}\;|_{\,_{\rm ref}}}$ or other 
observables, $R_O=\frac{~O}{O\;|_{\,_{\rm ref}}}$. The so determined 
observables, measured in units of their reference values, are universal
and can be compared at different $\beta$-values and for different actions.
\begin{figure}[t]
\vspace{-0.5cm}
\begin{center}

\ \ \ \ \ \includegraphics[angle=270,width=0.86\linewidth]{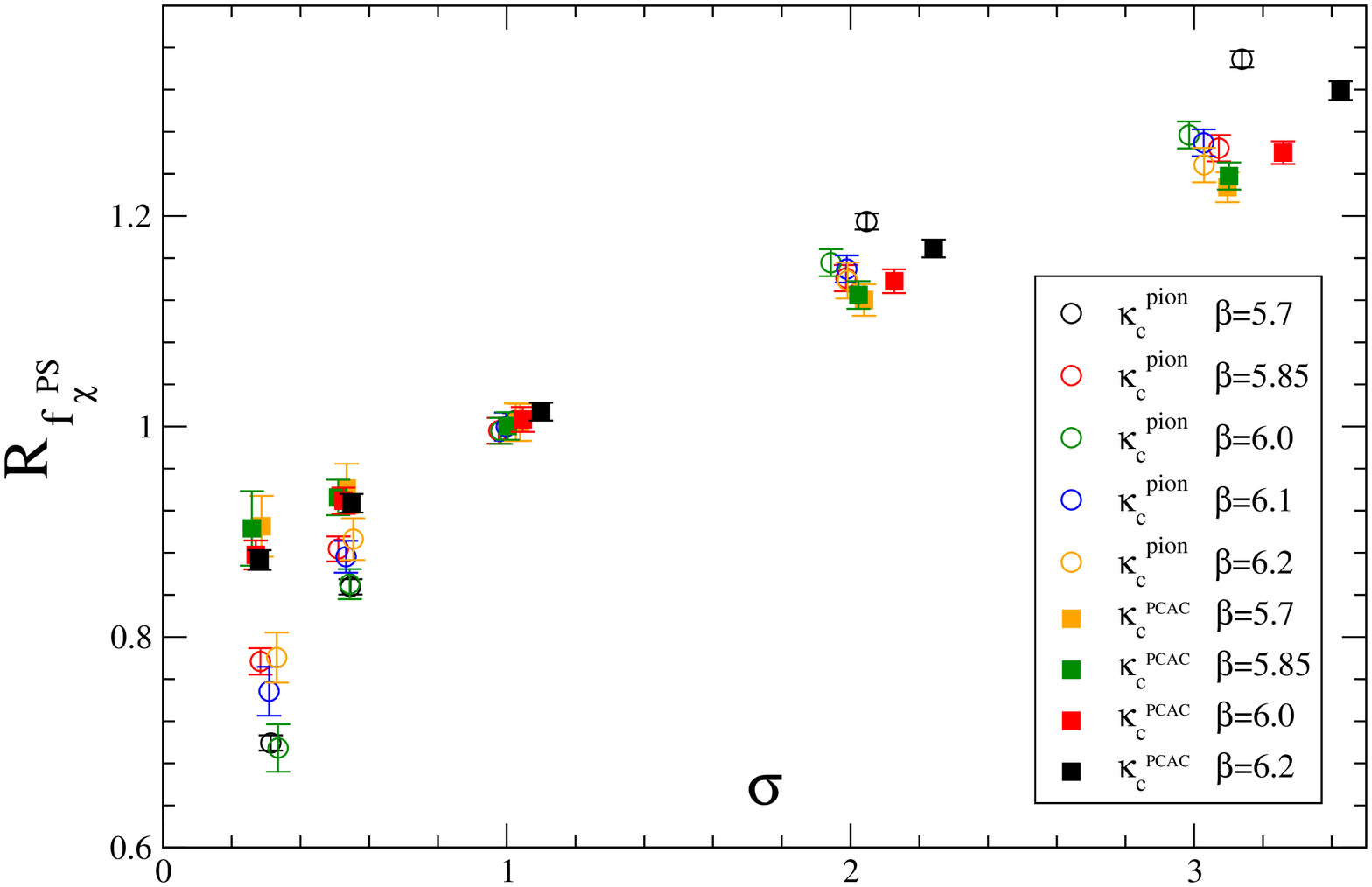}

\vspace*{-1.2cm}

\includegraphics[width=0.7\linewidth]{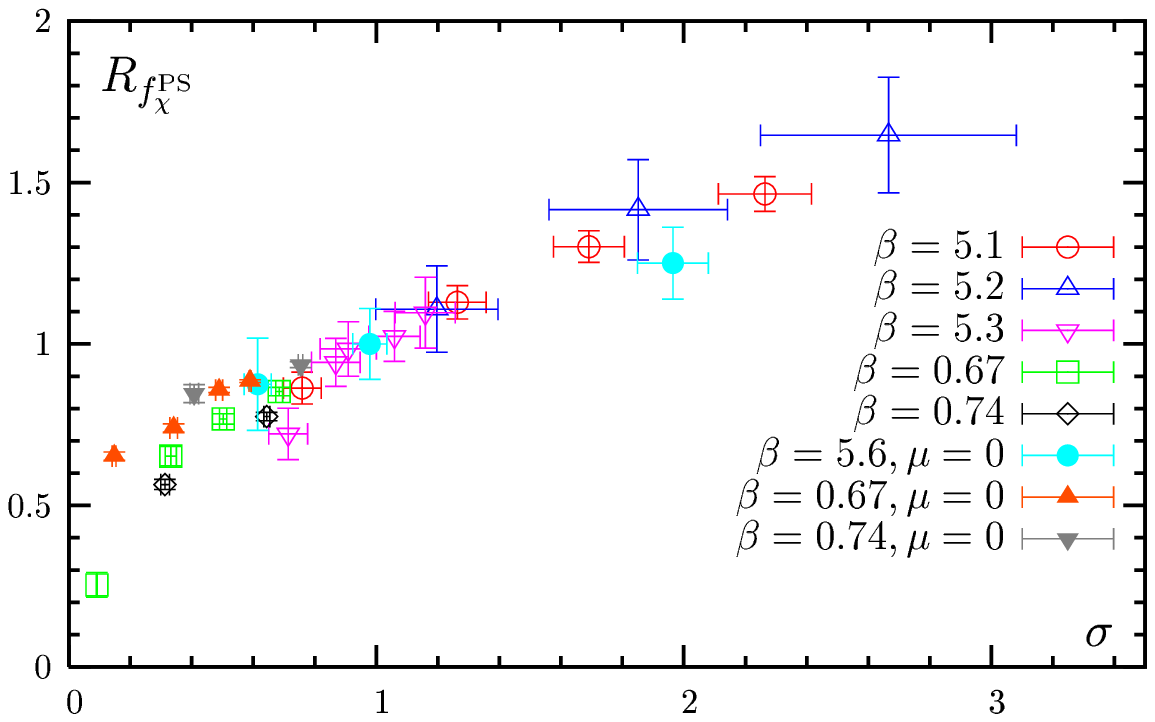}

\end{center}
\vspace{-0.7cm}
\caption{Scaling ratios $R_{f^{PS}_\chi}$ in the quenched approximation~(upper panel)
and for dynamical fermions~(lower panel). The lower panel also includes results from
the DBW2 action at $\beta=0.67$ and $\beta=0.74$ (see also sec.~4).
\label{fig:ratio}}
\end{figure}

In fig.~\ref{fig:ratio}~(lower panel) we show as an example the scaling ratio
for our dynamical fermion data of the Wilson plaquette and the DBW2 
gauge actions. We observe no scaling violations for the data from 
the Wilson plaquette gauge action at $\beta=5.1-5.3$ within the (large) statistical errors. 
Moreover, we see agreement with data at $\beta=5.6$ for pure Wilson fermions
without twisted mass, produced for the algorithmic study of an HMC variant
\cite{Urbach:2005ji,Urbach:proc}. For the data from the DBW2 gauge action
at smaller masses we observe instead a slight difference in the scaling
behavior of the $\mu=0$ and $\mu\ne0$ results. This might indicate that
such a difference could be detected for the Wilson plaquette gauge action only at
higher beta values, where smaller quark masses can be reached.
Similar plots were also obtained for the scaling ratios 
$R_{m_{\pi}^2}$ and $m_{\pi}/m_\rho$. 
For comparison, we also show in fig.~\ref{fig:ratio}~(upper panel) an example of the ratio
$R_{f^{PS}_\chi}$ for quenched data at full twist, $\kappa=\kappa_c$~\cite{Jansen:2005kk}.
Here we observe scaling violations for our coarsest lattice at $\beta=5.7$ and a difference
in the scaling behavior at small masses for different 
discretizations of twisted mass fermions, realized by using
two definitions of the critical mass ($\kappa_c^{pion}$ and $\kappa_c^{PCAC}$, 
see ref.~\cite{Jansen:2005kk}). 
This might provide a warning that at small values of the quark mass
the lattice artefacts could be significant and that scaling violations 
might show up. 

\section{DBW2 action: analysis of twist angle and  physical quantities}
\label{sec:omega_ren}

In this section we analyze several interesting quantities
in the context of simulations of $N_f=2$
flavors of twisted mass fermions with the DBW2 gauge action. 
This particular setup was studied 
in~\cite{Farchioni:2004fs} on a $12^3\cdot 24$ lattice at $\beta=0.67$
with spatial extension $L\simeq 2$~fm,
lattice spacing $a\simeq 0.18$~fm, $r_0\mu\simeq 0.03$: the minimal 
pion mass found here was $m_\pi\simeq 360$ MeV. Signs of metastabilities
are hardly detectable 
in this situation, which is a clear improvement compared to the 
Wilson plaquette gauge action, cf.~sec.~3. In this section  
we also consider data from $16^3\cdot 32$ lattices at a finer 
value of the lattice spacing, $\beta=0.74$, keeping  approximately 
the same volume and value of the twisted quark mass as in the 
aforementioned simulations.

\subsection{Twist angle}

\begin{figure}[htb]
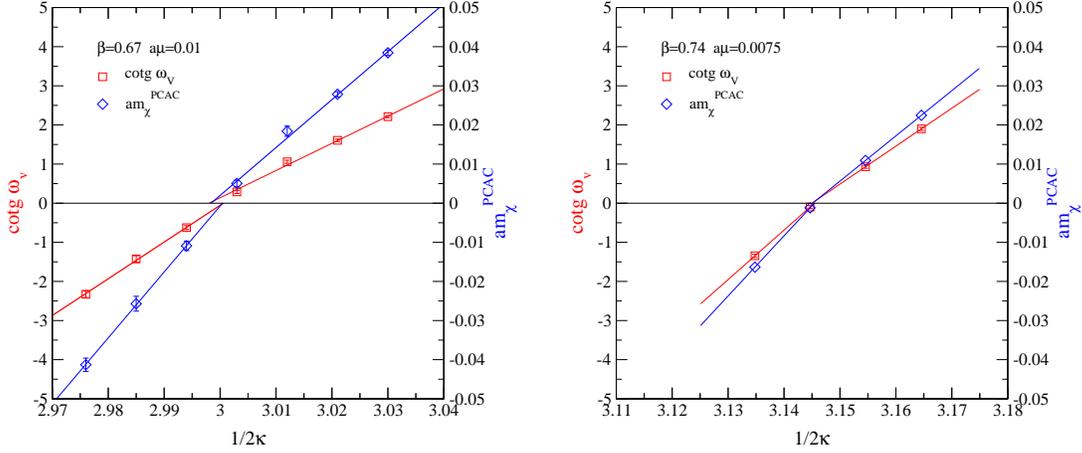

\vspace{+1cm}
\begin{center}
\epsfig{file=full_tw_extra_b067.eps,angle=0,width=0.45\linewidth}
\hspace{0.5cm}
\epsfig{file=full_tw_extra_b074.eps,angle=0,width=0.45\linewidth}
\end{center}
\vspace{-0.7cm}
\caption{Determination of the critical hopping parameter 
$\kappa_{c}$ corresponding to full twist  
by parity-restoration and by extrapolating to zero the untwisted PCAC quark mass $m_\chi^{PCAC}$.
The small discrepancy observed at $\beta=0.67$~(left panel) 
between extrapolations from positive and negative
quark masses is probably a residual effect of the first order phase transition (crossover).
For $\beta=0.74$~(right panel), extrapolations
from both sides give consistent results.
\label{fig:ctgom_mpcac}}
\end{figure}

The twist angle $\omega$ defines the chiral rotation relating 
twisted mass QCD to 
ordinary QCD. In the case of the chiral currents the rotation reads
(considering only charged currents, $a=1,2$):
\begin{eqnarray}\label{eq:phys_v}
 \hat{V}^a_{x\mu} &=& Z_V V^a_{x\mu}\,\cos\omega\, + 
\epsilon_{ab} \, Z_A A^b_{x\mu}\,\sin\omega \ ,
\\[0.5em]\label{eq:phys_a}
 \hat{A}^a_{x\mu} &=& Z_A A^a_{x\mu}\,\cos\omega\, +
\epsilon_{ab} \, Z_V V^b_{x\mu}\,\sin\omega\,
\end{eqnarray}
where the hatted currents on the l.h.s. denote the chiral 
currents of QCD (physical currents), while the currents on the r.h.s. are the
corresponding bilinears of the quark-field in the twisted ($\chi$-) basis. Note that
the renormalization constants of these bilinears, $Z_V$ and $Z_A$,
are involved. For a given choice of the lattice parameters, 
the twist angle $\omega$ is determined by requiring parity conservation
for matrix elements of the physical 
currents~\cite{Farchioni:2004ma,Farchioni:2004fs}. 
Since unknown renormalization constants 
are involved, {\em two} conditions are required, our choice being:
\begin{eqnarray}\label{eq:cond_v}
\sum_{\vec{x}} \langle \hat{V}^+_{x0}\, P^-_y\rangle&=& 0\ ,
\\
\label{eq:cond_a}
\sum_{\vec{x},i} \langle \hat{A}^+_{xi}\, \hat{V}^-_{yi}\rangle&=& 0\ . 
\end{eqnarray}
The solution of eqs.~(\ref{eq:cond_v}) and (\ref{eq:cond_a}) with
eqs.~(\ref{eq:phys_v}) and (\ref{eq:phys_a}) gives a direct 
determination of the twist angle $\omega$ and of the ratio $Z_A/Z_V$ from lattice data, 
see~\cite{Farchioni:2004fs} for details. Eq.~(\ref{eq:cond_v}) implies e.g.
\be\label{eq:cotomv}
\cot{\omega}_V\equiv \frac{Z_V}{Z_A} \cot\omega =
-i\, \frac{  \sum_{\vec{x}}\langle A^+_{x0}\, P^-_y \rangle} 
{  \sum_{\vec{x}}\langle V^+_{x0}\, P^-_y \rangle}\ .
\ee
In particular at full twist where $\omega=\pi/2$ the condition reads
$\displaystyle \sum_{\vec{x}}\langle A^+_{x0}\, P^-_y \rangle = 0$ or $\cot{\omega}_V=0$.
The full twist situation can be also obtained by requiring the 
vanishing of the PCAC quark mass $m_\chi^{PCAC}$. One can easily see that 
using the PCAC quark mass as a criterion for full twist just amounts to
a different definition of the twist angle: the current is replaced by its divergence
in~(\ref{eq:cond_v}). So the two criteria are equivalent and differ (at most) by $O(a)$ effects.
The numerical equivalence of the two criteria 
for full twist is illustrated by fig.~\ref{fig:ctgom_mpcac}.

\subsection{Physical quark mass and pion decay constant}

\begin{figure}[htb]
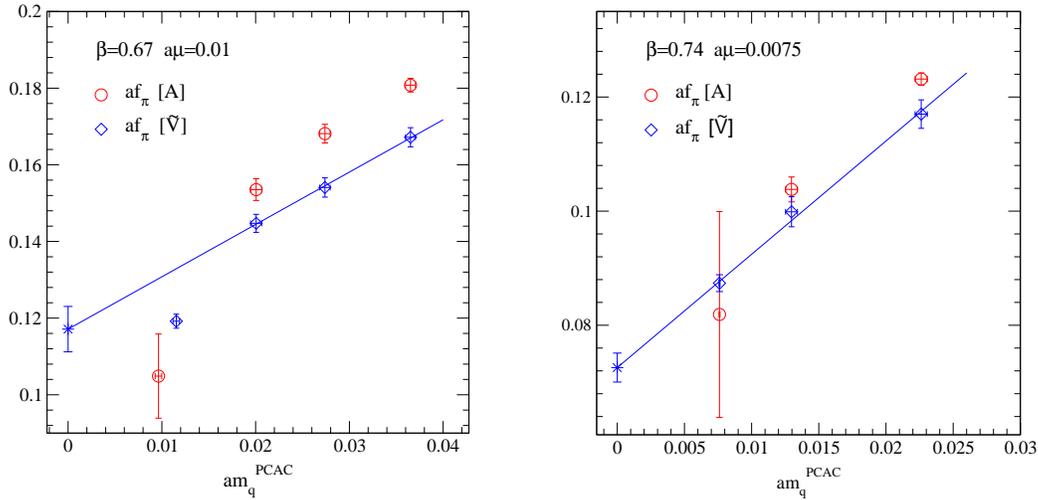

\vspace{+1cm}
\begin{center}
\epsfig{file=fpi_vs_mq_dbw2_b067.eps,angle=0,width=0.407\linewidth}
\hspace{1cm}
\epsfig{file=fpi_vs_mq_dbw2_b074.eps,angle=0,width=0.42\linewidth}
\end{center}
\vspace{-0.7cm}
\caption{The pion decay constant $f_\pi$ as a function of the PCAC quark mass 
$m_q^{PCAC}$. 
\label{fig:fpi_vs_mq}}
\end{figure}

The knowledge of the twist angle is necessary for the determination
of physical quantities like the quark mass and the pion decay constant.
The {\em physical} PCAC quark mass $m_q^{PCAC}$ can be obtained from the Ward identity for the 
physical axialvector current:
\be\label{eq:phys_ward}
\langle\nabla^*_\mu\hat A^{+}_{x\mu} P^{-}_y\rangle = 2 m_q^{PCAC}
\langle P^{+}_{x} P^{-}_y\rangle \ .
\ee
From eqs.~(\ref{eq:phys_v}) and (\ref{eq:phys_a}) follows 
\be\label{eqX:28}
 \hat A^a_{x\mu}  = \hat V^a_{x\mu}\,\cot\omega\, +
\epsilon_{ab} \, \tilde{V}^b_{x\mu} \,(\sin\omega)^{-1}
\ee
where the conserved vector current of the $\chi$-fields $\tilde{V}^b_{x\mu}$ has been now considered,
for which $Z_V=1$. Inserting the above result in the Ward identity~(\ref{eq:phys_ward})
and using parity restoration for the physical currents one obtains:
\be\label{eq:mpcac_phys}
 m_q^{PCAC} = -i
(2\sin\omega)^{-1}
\frac{\langle\nabla^*_\mu \tilde V^{+}_{x\mu} P^{-}_y\rangle} 
     {\langle P^{+}_x P^{-}_y\rangle}
= Z_A(\cos\omega)^{-1}m_\chi^{PCAC}\ .
\ee
Analogously, for the physical pion decay constant $f_{\pi}$ we use
\be\label{eq:fpi_phys}
f_{\pi}=m_\pi^{-1} \langle 0|\hat A^+_0(0)|\pi^+\rangle=
-i(m_\pi \sin\omega)^{-1} \langle 0|\tilde V^+_0(0)|\pi^+\rangle\ .
\ee
In fig.~\ref{fig:fpi_vs_mq} the pion decay constant is plotted as a function of the 
quark mass (the simulation points for negative quark
masses are not included). The figures also show the determination of $f_{\pi}$ 
by the axialvector current $A^a_{x\mu}$: a formula similar 
to eq.~(\ref{eq:fpi_phys}) applies in this case, where however the factor 
$1/\sin\omega$ is replaced by  $1/\cos\omega$. In the interesting
region near full twist this factor introduces the large fluctuations in the estimate of $f_{\pi}$
observed in the data. Moreover in the case of $A^a_{x\mu}$
the decay constant has not yet the right normalization of the continuum:  
a $Z_A$ factor is missing. In the case of the conserved vector current on the contrary
the lattice determination of $f_{\pi}$ has automatically the correct 
normalization~\cite{Frezzotti:2001du,Jansen:2003ir}. If we exclude the lightest point at $\beta=0.67$, which 
is likely to be under the influence of residual metastability effects, 
$f_{\pi}$ seems to 
be characterized by a linear dependence upon the quark mass. On the basis of this observation
we try a simple linear extrapolation to the chiral limit $m_q^{PCAC}=0$, 
see table~\ref{tab_chiral_extra} for the numerical results.
(Of course deviations from  this linear behavior could be present for lighter quark masses, 
where chiral logarithms might play a role.)
We see that the two $\beta$-values give compatible values for 
$f_{\pi} r_0$, suggesting that there are no large scaling violations
for this action. These values, obtained with $N_f=2$, 
are also near to the physical value $f_{\pi} r_0=0.33$.
\begin{table}
\centering
\begin{center}
\parbox{0.8\linewidth}{\caption{\label{tab_chiral_extra}
Chiral extrapolation ($m_q^{PCAC}=0$) of the Sommer scale parameter $r_0$ and pion decay constant
$f_\pi$; the scale independent combination $f_\pi r_0$ is also reported.
Only data with positive twisted quark masses have been used for the extrapolations, 
with the exception of the point near full twist at $\mu=7.5\, \cdot 10^{-3}$.
}}
\end{center}
\begin{center}
 \begin{tabular}{lllllll}
\hline
\hline
     \multicolumn{1}{c}{$\beta$} &
     \multicolumn{1}{c}{$\mu$} & 
     \multicolumn{1}{c}{$r_0/a$} & 
     \multicolumn{1}{c}{$a$ (fm)} & 
     \multicolumn{1}{c}{$a\, f_\pi$}&     
     \multicolumn{1}{c}{$f_\pi\,r_0$} 
    \\
\hline
\hline
0.67 & $1.0\, \cdot 10^{-2}$  & 2.680(68) & 0.1866(72)  & 0.1171(59)  & 0.320(10)    \\
0.74 & $7.5\, \cdot 10^{-3}$  & 4.11(13)  & 0.1216(39)  & 0.0726(25)  & 0.309(15)    \\
\hline
\hline
\end{tabular}
\end{center}
\end{table}

\subsection{Renormalization constants}

The renormalization constant of the vector current $Z_V$ can be determined on the basis of the 
non-renormalization property of the conserved current $\tilde V_{x\mu}$~\cite{Maiani:1986yj}. 
In the case of twisted mass QCD we use
\be\label{eq:zv}
Z^{latt}_V=\frac{\langle 0|\tilde V_0^+|\pi^+\rangle}{\langle 0|V_0^+|\pi^+\rangle}\ .
\ee
Note that in twisted mass QCD the time-component of the 
vector current couples the vacuum to the pseudoscalar meson: 
in the most interesting region near full twist 
this coupling is (in our twisted basis) maximal. 
In lattice QCD with ordinary Wilson fermions the analogous procedure 
has to rely on the noisier matrix element of the vector particle.
Eq.~(\ref{eq:zv}) represents a mass dependent prescription.
We obtain a mass independent determination of $Z_V$ by extrapolating  $Z^{latt}_V$ to full twist. 
In this situation the theory is $O(a)$ improved and $Z^{latt}_V$ has only $O(a^2)$ 
lattice deviations. (In particular residual $O((\mu a)^2)$ terms can be eliminated 
by a $\mu\rightarrow 0$ extrapolation.)
The results are reported in table~\ref{tab_zfac_ft}; 
the renormalization of the axialvector 
current $Z_A$ is also reported: this is obtained by combining  the direct determination
of $Z_V$~(\ref{eq:zv}) and the determination of $Z_A/Z_V$ from the parity restoration (see above)
after extrapolation to full twist. Observe that the 
conditions~(\ref{eq:cond_v}),~(\ref{eq:cond_a}) hold in general up to $O(a)$ 
parity violations: imposing (as we do) exact parity restoration for the considered m.e.
leads to a mass dependent determination of the ratio $Z_A/Z_V$.
\begin{table}
\centering
\begin{center}
\parbox{0.8\linewidth}{\caption{\label{tab_zfac_ft}
The renormalization constants $Z_V$, $Z_A$ at full twist (extrapolating 
from positive quark masses) with comparison to 1-loop perturbative estimates (PT) 
and tadpole-improved perturbative 
estimates (TI)\cite{Perlt}.
}}
\end{center}
\begin{center}
\begin{tabular}{llclll}
\hline
\hline
     \multicolumn{1}{c}{$\beta$} &
     \multicolumn{1}{c}{$\mu$} &
     \multicolumn{1}{c}{Op.} &
     \multicolumn{1}{c}{$Z$} &
     \multicolumn{1}{c}{$Z$(PT)}&
     \multicolumn{1}{c}{$Z$(TI)}
    \\
\hline
\hline
0.67 & $1.0\, \cdot 10^{-2}$ & $V$ & 0.5650(11)  &0.6089 &0.6531              \\
0.74 & $7.5\, \cdot 10^{-3}$ & $V$ & 0.6217(23)  &0.6459 &0.6892              \\
0.67 & $1.0\, \cdot 10^{-2}$ & $A$ & 0.952(30)   &0.7219 &0.7176             \\
0.74 & $7.5\, \cdot 10^{-3}$ & $A$ & 0.944(74)   &0.7482 &0.7735             \\
\hline
\hline
\end{tabular}
\end{center}
\end{table}

\section{Tree-level Symanzik improved gauge action}

In this section we will now concentrate on the tree-level Symanzik improved
(tlSym) gauge action for which the coefficient $c_1=-1/12$.  The reasons for
the choice of this action are manifold -- firstly, the action is designed to
show a good and smooth behavior in the perturbative regime; secondly, in many
quenched scaling studies the action has proved to behave well also in the
non-perturbative regime; and thirdly, the additional coefficient $c_1$ is
rather small in absolute value as compared to other improved actions like the
DBW2.
We have simulated several values of $\kappa$ close to $\kappa_c$, i.e.~close
to maximal twist, at three different values of $\beta$ using a HMC algorithm
variant with mass preconditioning and multiple time scale 
integration~\cite{Urbach:2005ji,Urbach:proc}.

\begin{figure}[htb]
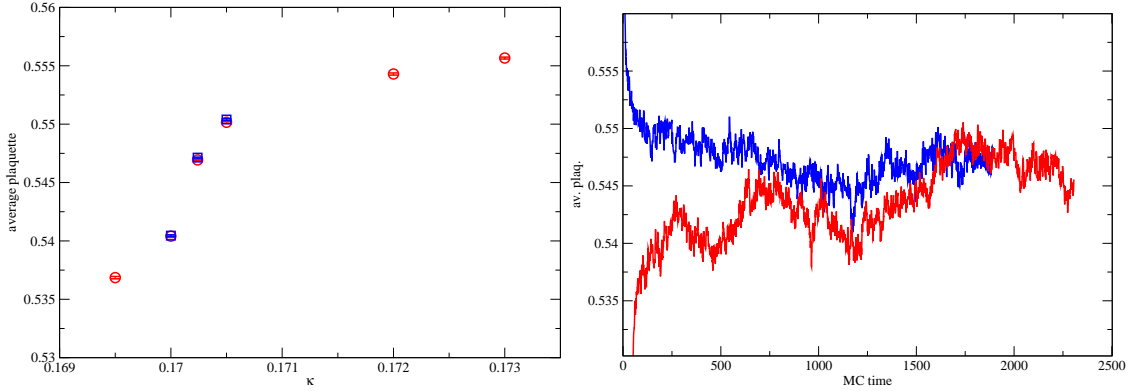

\vspace{-0.0cm}
\begin{center}
\includegraphics[width=0.49\linewidth]{av_plaq_tlSym_b3.65_m0.01_L12T24.eps}
\includegraphics[width=0.49\linewidth]{L12T24_b3.65_m0.01_plaq_mc_history.eps}
\end{center}
\vspace{-0.7cm}
\caption{Left panel: Average plaquette value 
vs.~$\kappa$ at $\beta=3.65, r_0 \mu=0.038$
  on a $12^3\cdot 24$ lattice from hot (red circles) and cold starts (blue squares).
Right panel: 
Plaquette MC time history for two runs at $\beta=3.65, r_0 \mu=0.038, \kappa=0.17024$
  on a $12^3\cdot 24$ lattice starting from hot (red line coming from below) and cold configuration
  (blue line coming form above).
\label{fig:av_plaq_tlSym_b3.65_m0.01_L12T24}}
\end{figure}

In order to check for a possible phase transition and corresponding
metastabilities we measure the average plaquette value as a function of the
hopping parameter $\kappa$ on runs with hot (disordered) and cold (ordered) starting 
configurations. Since the metastability, if any, will show up around $\kappa_c$
we concentrate our hot and cold runs on $\kappa$-values closest to $\kappa_c$
only. Fig.~\ref{fig:av_plaq_tlSym_b3.65_m0.01_L12T24}~(left) shows the results for
the average plaquette value for our coarsest lattice at $\beta=3.65, 12^3\cdot 24$
and $r_0 \mu=0.038$. Here and in the rest of the section the red circles and blue squares
represent the results from the hot and cold starts, respectively.
From fig.~\ref{fig:av_plaq_tlSym_b3.65_m0.01_L12T24}~(left) it is clear that there
seems to be no metastability as far as the plaquette value is concerned, but
the existence of a phase transition at this parameter set can of course not be
excluded. Indeed, the rise of the plaquette value around
$\kappa_c\simeq0.17025$ points toward the fact that we are at least in the
vicinity of a phase transition and we conclude that a lower value of $\mu$
would hardly be possible to simulate in practice. The vicinity of the phase
transition is also reflected in the MC time histories of the plaquette value.
We observe a rather strong critical slowing down leading both to a very slow
thermalization and large fluctuations of the plaquette value over several
hundreds of trajectories. It is evident from the MC time history shown in
fig.~\ref{fig:av_plaq_tlSym_b3.65_m0.01_L12T24}~(right) that any statement about
the plaquette value itself as well as any error estimate are very difficult to
make and should hence be taken with care. The figure also illustrates the fact
that simulations close to or at $\kappa_c$ (i.e.~at maximal twist angle) are
rather difficult, at least at such large values of the lattice spacing
($a\simeq 0.13$~fm) and values of $\mu$ as chosen here.

\begin{figure}[htb]
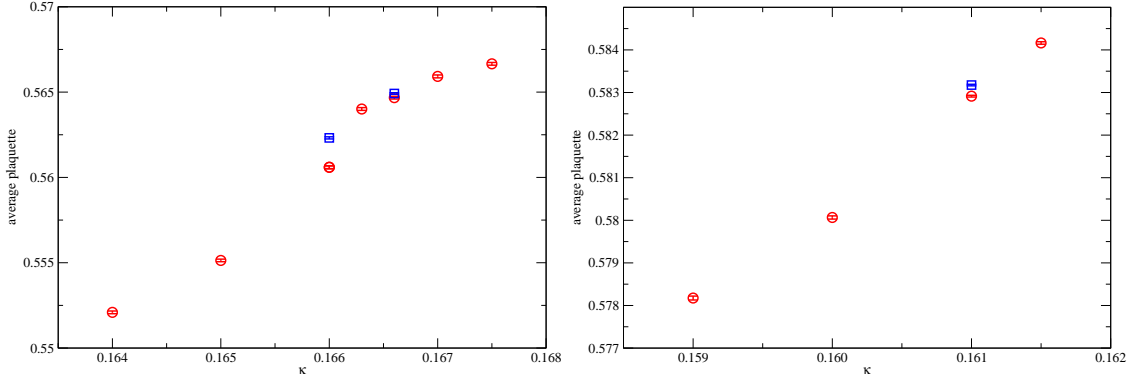

\vspace{-0.0cm}
\begin{center}
\includegraphics[width=0.49\linewidth]{av_plaq_tlSym_b3.75_m0.005_L12T24.eps}
\includegraphics[width=0.49\linewidth]{av_plaq_tlSym_b3.90_m0.0075_L16T32.eps}
\end{center}
\vspace{-0.7cm}
\caption{Left panel: Average plaquette value vs.~$\kappa$ at $\beta=3.75, r_0 \mu=0.020$
  on a $12^3\cdot 24$ lattice from hot (red circles) and cold starts (blue squares).
Right panel: Average plaquette value vs.~$\kappa$ at $\beta=3.90, r_0 \mu=0.041$
  on a $16^3\cdot 32$ lattice from hot (red circles) and cold starts (blue squares).
\label{fig:av_plaq_tlSym_b3.75_m0.005_L12T24}}
\end{figure}

Let us now move to our simulations at $\beta=3.75$. 
Fig.~\ref{fig:av_plaq_tlSym_b3.75_m0.005_L12T24}~(left) shows the average plaquette value
vs.~$\kappa$ at $\beta=3.75, r_0 \mu=0.020$ on a $12^3\cdot 24$ lattice from hot
(red circles) and cold starts (blue squares). Compared to 
fig.~\ref{fig:av_plaq_tlSym_b3.65_m0.01_L12T24} we are now at finer lattice spacing
($r_0/a=3.76(9)$ at $\beta=3.65$ vs.~$r_0/a=4.1(2)$ at $\beta=3.75$), 
but on the other hand
also our value of the twisted mass is halved, so we do not necessarily expect
the situation to be better.
Indeed, we find that at the value of $\kappa=0.1660$ that is closest to
$\kappa_c$ the average plaquette value from the hot and the cold start do not
match each other, which is pointing to the existence of a phase transition and
a corresponding metastability at this set of parameters. On the other hand,
considering the large fluctuations of the plaquette similar to the ones in
fig.~\ref{fig:av_plaq_tlSym_b3.65_m0.01_L12T24}, the runs here are
presumably not long enough to allow reliable estimates and the metastability
might disappear as the MC time history becomes longer. Nevertheless, the
strong rise of the plaquette around $\kappa_c$ points toward the fact that we
are at least close to a transition, even though the rise appears to be much
weaker than at $\beta=3.65$.

The situation at $\beta=3.90$ is illustrated in 
fig.~\ref{fig:av_plaq_tlSym_b3.75_m0.005_L12T24}~(right) where $r_0 \mu=0.041$
corresponds again to our larger value of $\mu$. Here we observe a very
smooth dependence of the plaquette value on $\kappa$ and there seems to be no
trace of a cross-over or even a nearby phase transition. (We attribute the fact
that the plaquette value at $\kappa=0.1610 \simeq \kappa_c$ from the hot and
cold start do not match exactly to the shortness of our runs
which prevents an accurate determination of the average plaquette value.)

\begin{figure}[htb]
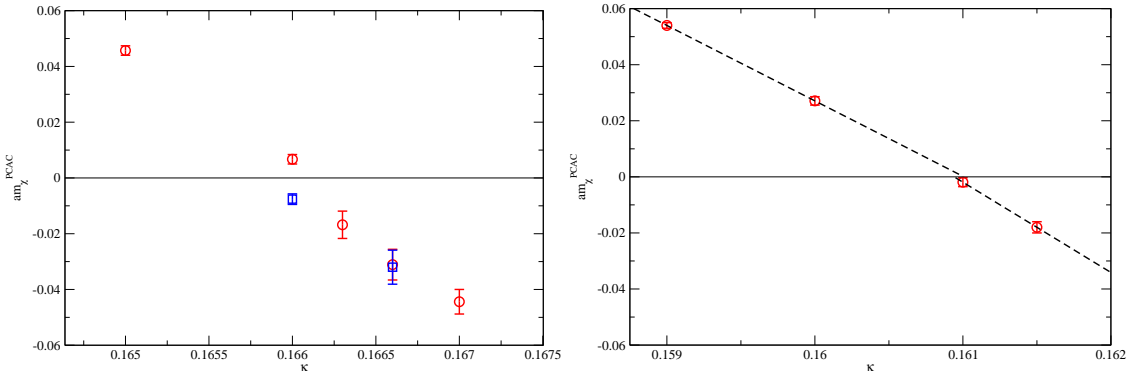

\vspace{-0.0cm}
\begin{center}
\includegraphics[width=0.49\linewidth]{tlSym_L12T24_b3.75_m0.005_mPCAC_vs_kappa.eps}
\includegraphics[width=0.49\linewidth]{tlSym_L16T32_b3.90_m0.0075_mPCAC_vs_kappa.eps}
\end{center}
\vspace{-0.7cm}
\caption{Untwisted PCAC quark mass $m_\chi^{PCAC}$ vs.~$\kappa$ from hot (red circles)
  and cold starts (blue squares). At $\beta=3.75, r_0 \mu=0.020$
  on a $12^3\cdot 24$~(left plot) and $\beta=3.90, r_0
  \mu=0.041$ on a $16^3\cdot 32$ lattice~(right plot).
\label{fig:tlSym_mPCAC_vs_kappa}}
\end{figure}

Next we consider the untwisted PCAC quark mass $m_\chi^{PCAC}$, cf.~eq.~(\ref{eqpcacmass}),
as a function of $\kappa$ for $\beta=3.75$ and $3.90$ in 
fig.~\ref{fig:tlSym_mPCAC_vs_kappa}.
For $\beta=3.75$~(left) we find that the mismatch between the plaquette
value from the hot and the cold start at $\kappa=0.1660 \simeq \kappa_c$ is
also reflected in the PCAC quark mass, although we expect that this mismatch
will eventually go away for long enough runs as discussed before.
Nevertheless, even then it seems unlikely that extrapolations to
$m_\chi^{PCAC}=0$ from $\kappa$-values above and below $\kappa_c$ will
coincide. This is in contrast to the situation at $\beta=3.90$~(right plot)
where we observe a smooth dependence of $m_\chi^{PCAC}$ on $\kappa$ throughout
a wide range of $\kappa$-values.  Indeed, extrapolations to $m_\chi^{PCAC}=0$
from $\kappa$-values above and below $\kappa_c$ nicely coincide while    
the slopes are different (in consistency with expectations from
$\chi$PT~\cite{Munster:2004am,Scorzato:2004da,Sharpe:2004ny,Sharpe:2004ps,Aoki:2004ta}).

\begin{figure}[htb]
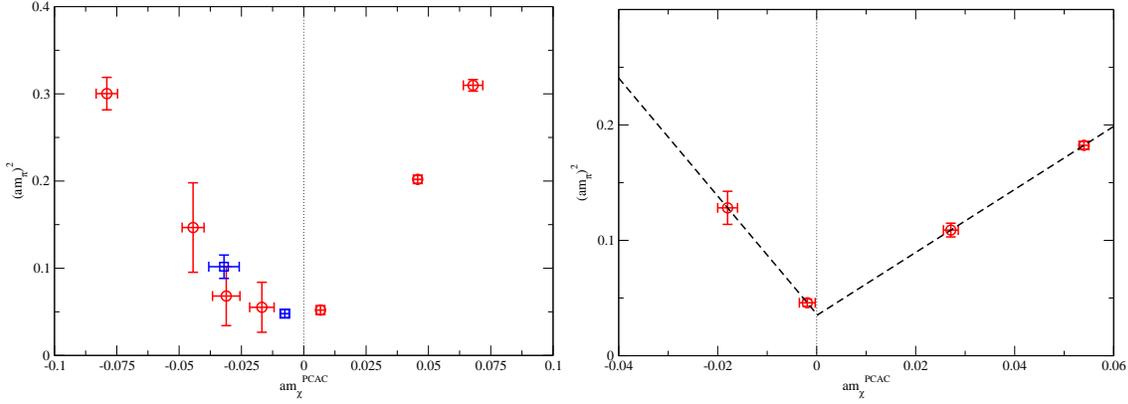

\vspace{-0.0cm}
\begin{center}
\includegraphics[width=0.49\linewidth]{tlSym_L12T24_b3.75_m0.005_mPi_vs_mPCAC.eps}
\includegraphics[width=0.49\linewidth]{tlSym_L16T32_b3.90_m0.0075_mPi_vs_mPCAC.eps}
\end{center}
\vspace{-0.7cm}
\caption{Pion mass squared $m_\pi^2$ vs.~PCAC quark mass $m\chi^\text{PCAC}$ from hot (red circles)
  and cold starts (blue squares). At $\beta=3.75, r_0 \mu=0.020$
  on a $12^3\cdot 24$~(left plot) and $\beta=3.90, r_0
  \mu=0.041$ on a $16^3\cdot 32$ lattice~(right plot).
\label{fig:tlSym_m_pi_vs_mPCAC}}
\end{figure}

Let us now turn to the pion mass squared $m_\pi^2$ as a function of
the PCAC quark mass $m_\chi^{PCAC}$ which we plot in 
fig.~\ref{fig:tlSym_m_pi_vs_mPCAC} for $\beta=3.75, r_0 \mu=0.020, 12^3\cdot 24$ on the
left and for $\beta=3.90, r_0 \mu=0.041, 16^3\cdot 32$ on the right.
At $\beta=3.75$ the pion mass that is realized at $r_0\mu=0.020$ is $m_\pi =
353(22)$~MeV for the hot start and $m_\pi = 382(33)$~MeV for the cold start, so
$m_\pi \simeq 370$~MeV seems to be a good estimate for the minimal pion mass
that can be reached at this parameter set. 
For additional simulations at the same $\beta$ but on a larger volume
$16^3\cdot 32$ at $\kappa \simeq  \kappa_c$ and $r_0 \mu = 0.039$ we obtain
$m_\pi=407(16)$~MeV.
Finally, at $\beta=3.90$ the pion mass we reach at $r_0\mu=0.041$ is $m_\pi =
453(32)$~MeV, while an additional simulation at $\kappa\simeq\kappa_c$
and $r_0\mu=0.020$ gives $m_\pi = 279(37)$~MeV.

\section{Chiral fits}

The extension of $\chi$PT to the case of adding a twisted mass term 
was considered in refs.~\cite{Munster:2003ba,Munster:2004am,Scorzato:2004da,Sharpe:2004ny,Sharpe:2004ps,Aoki:2004ta}.
For the low-energy constants in next-to-leading order
\cite{Rupak:2002sm,Bar:2003xq} we use the notation
$L_{54} = 2 L_4 + L_5$, $L_{86} = 2 L_6 + L_8$, 
$W_{54} = 2 W_4 + W_5$, $W_{86} = 2 W_6 + W_8$, 
$W = \frac{1}{2} (W_{86} - 2 L_{86})$,
$\widetilde{W} = \frac{1}{2} (W_{54} - L_{54})$
where the $L_i$ are the Gasser-Leutwyler coefficients and the  
$W_i$ are the corresponding coefficients of the $O(m_q a)$ corrections.
We consider the dependence of lattice quantities on the untwisted
PCAC quark mass $m_\chi^{PCAC}$, so we define $\chi'_{PCAC}
= 2 B_{0}\, m^{PCAC}_{\chi R}$ where we explicitly express that 
$m^{PCAC}_{\chi}$ is renormalized in some prescription: $m^{PCAC}_{\chi R}= Z_AZ_P^{-1} m^{PCAC}_{\chi}$.
We also define the combination containing the physical quark mass
$\bar{\chi}= 2 B_{0} \sqrt{(m^{PCAC}_{\chi R})^2 + \mu_R^2}$
with $\mu_R=Z_P^{-1}\mu$.
Defining $\rho = 2 W_0 a$, where $W_0$ is the low-energy scale for $O(a)$ 
breaking (analogous to $B_0$), $F_0$ the pion decay constant in the chiral limit, and 
$\Lambda$ the renormalization scale, 
$\chi$PT at next-to-leading order with $O(a)$ corrections gives:
\begin{eqnarray}
\label{mpi-chpt}
m_{\pi^\pm}^2 &=& \bar{\chi}
+ \frac{1}{32 \pi^2 F_0^2} \bar{\chi}^2
\ln \frac{\bar{\chi}}{\Lambda^2}
+ \frac{8}{F_0^2}
\{ (- L_{54} + 2 L_{86}) \bar{\chi}^2
+ 2 (W - \widetilde{W}) \rho\,
\chi'_{PCAC} \} \\
\label{fpi-chpt}
\frac{f_{\pi R}}{F_0} &=& 1
- \frac{1}{16 \pi^2 F_0^2} \bar{\chi} \ln \frac{\bar{\chi}}{\Lambda^2}
+ \frac{4}{F_0^2} \{ L_{54} \bar{\chi}
+ 2 \widetilde{W} \rho\,
\frac{\chi'_{PCAC}}{\bar{\chi}} \}\\
\label{gpi-chpt}
\frac{g_{\pi R}}{F_0 B_0} &=& 1
- \frac{1}{32 \pi^2 F_0^2}
\bar{\chi} \ln \frac{\bar{\chi}}{\Lambda^2}
+ \frac{4}{F_0^2} \{ (- L_{54} + 4 L_{86}) \bar{\chi}
+ ( 4 W - 2 \widetilde{W}) \rho\,
\frac{\chi'_{\rm\scriptscriptstyle PCAC}}{\bar{\chi}} \}\ .
\end{eqnarray}
We set $\Lambda=4\pi F_0$.

\begin{figure}[htb]
\vspace{-0.0cm}
\begin{center}
\epsfig{file=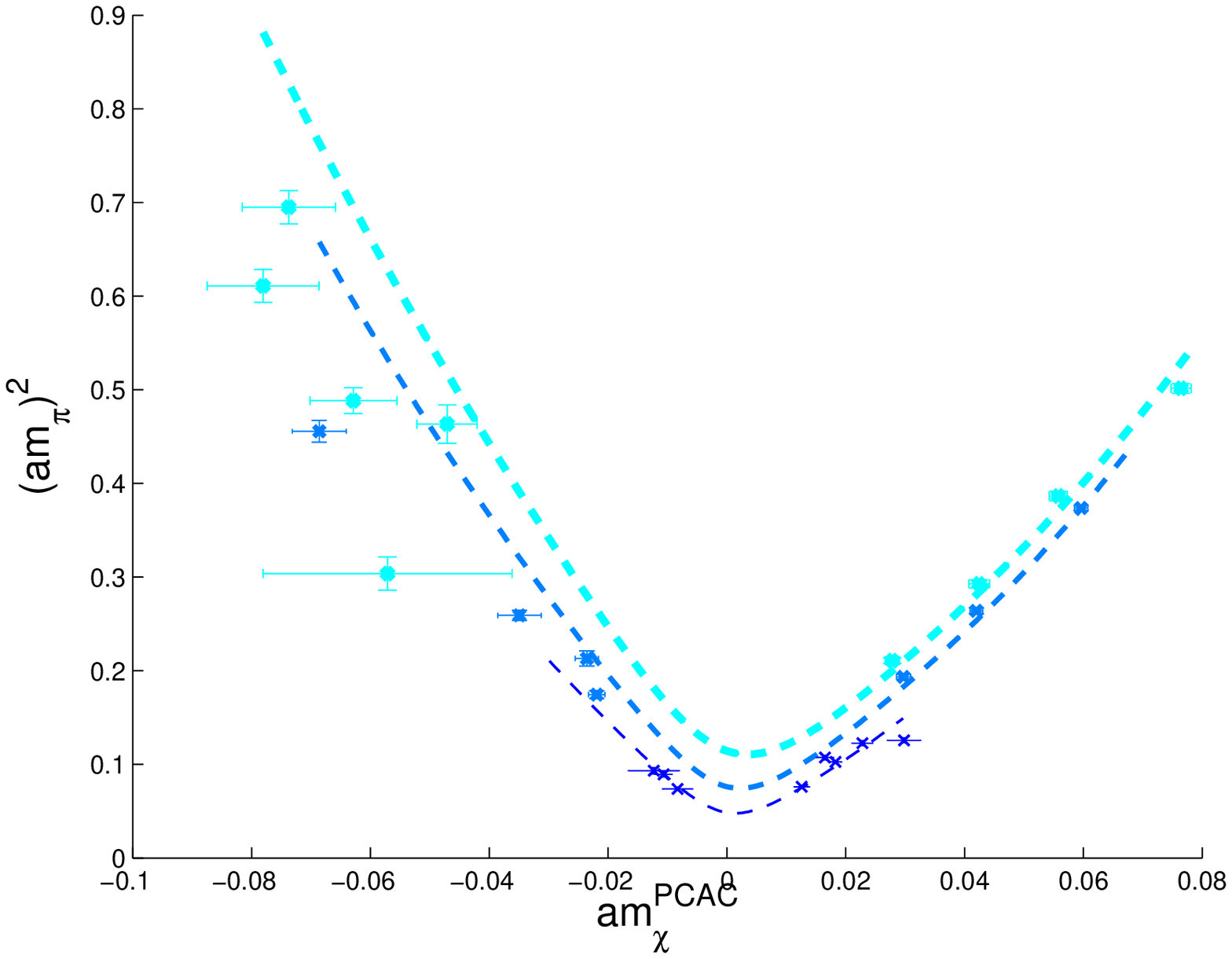,width=0.45\linewidth}
\epsfig{file=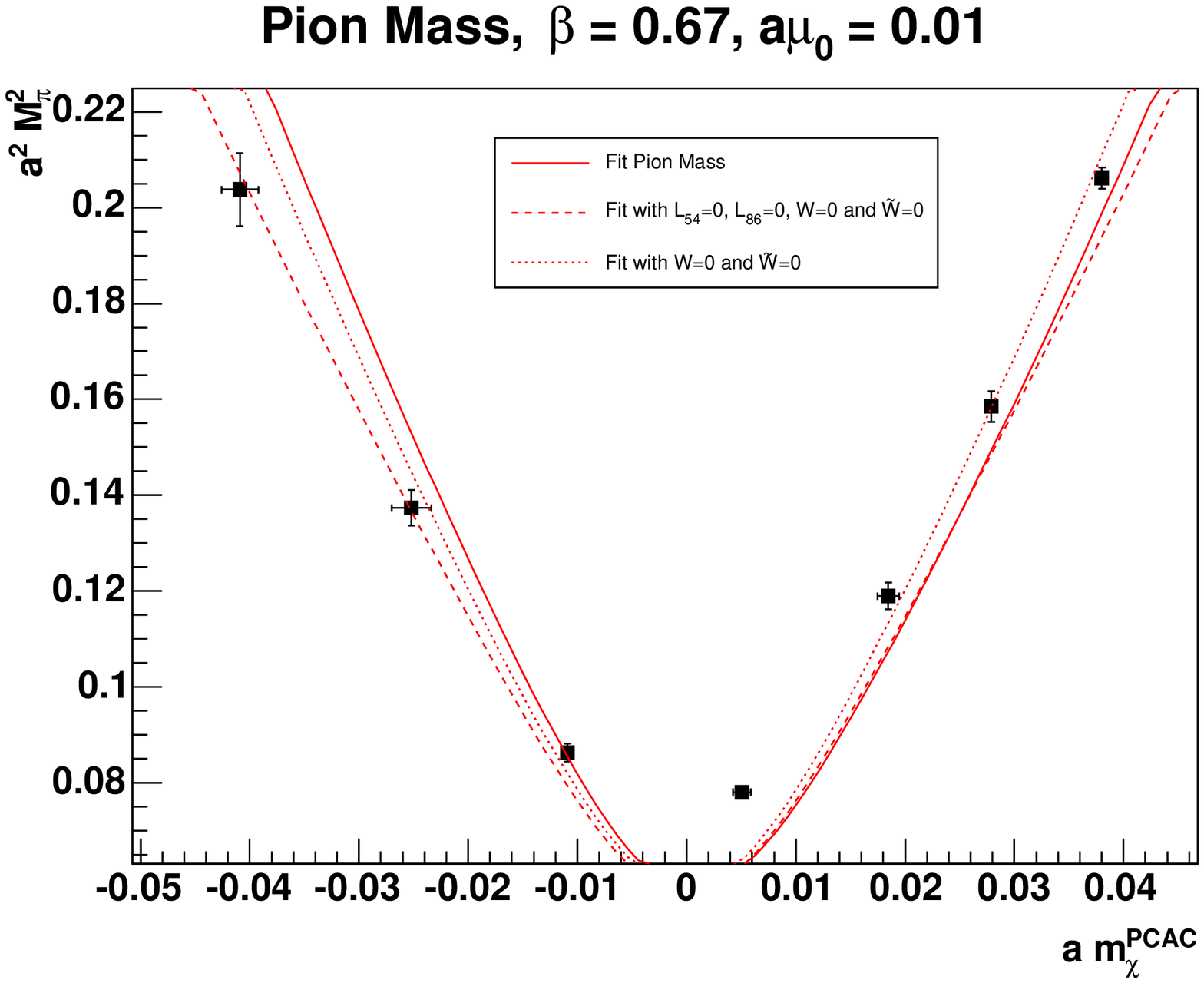,width=0.45\linewidth}
\end{center}
\vspace{-0.7cm}
\caption{Chiral fit of the pion mass: Wilson plaquette action~(left) and DBW2 action~(right). 
In the Wilson case a darker blue tone corresponds to a larger value of $\beta$.
\label{fig:chifit}}
\end{figure}

We performed a combined chiral fit of the quantities $m_{\pi}^2, f_\pi$ and
$g_{\pi}$ for our data from the Wilson plaquette and the DBW2 gauge action
(including three and two different $\beta$-values, respectively, see sec.~3 and 4)
in order to obtain estimates for the combinations of coefficients
$L_{86}$, $L_{54}$, $W$, $\tilde{W}$ and of the low-energy constants $B_0$ and  $F_0$. 
In fig.~\ref{fig:chifit} we show the result of this global chiral fit for the case of the pion mass,
for the Wilson plaquette action~(left) and the DBW2 action (right). In the former case 
three values of $\beta$ are included.

The general observation is that lattice $\chi$PT theory formulae reproduce rather well the
lattice data.
The $O(a)$ terms of the chiral expansion are in general not dominant
in the fits, which one can infer from the fact that the associated 
coeffiecients fluctuate with large deviations around zero.
This is probably due to the use of the PCAC quark mass which in past experiences
has shown to have reduced lattice artefacts compared to the $1/2\kappa$ definition of the bare 
quark mass. In fact fig.~\ref{fig:plaq}~(right panel) shows the presence of large lattice
artifacts.

The use of different setups for the fits (different sets of data-points and actions, independent
fit methodologies) allowed to test the stability of the results for the low-energy
constants. Combining all information, we get the following ranges (here we fix the lattice scale
 by using the chiral-extrapolated value of the Sommer scale $r_0$): $F_0=70-100$~MeV, $B_0/Z_P=3-6$~GeV,
$L_{54}=(0.8-1.8)\cdot10^{-3}$, $L_{86}=(0.5-1.0)\cdot10^{-3}$. As mentioned, the coefficients of the $O(a)$
terms could not be determined.

\section{$N_f=2+1+1$ flavors of twisted mass fermions}

In this section we consider simulations of QCD in the twisted mass setup
including the dynamics of the {\em strange} ($s$) and {\em charm} ($c$) flavors. 
The $u$ and $d$ quarks, which are degenerate, are still described by 
the action~(\ref{eq:ferm_action}).

\subsection{Split doublet}

The mass-splitting in the charm-strange ($cs$) sector is realized by adding  
an extra mass-term~\cite{Frezzotti:2003xj} in the action for a doublet 
of degenerate twisted mass fermions:
\be\label{eq:plit_doub_twba}
S^{cs}_q = \sum_x \left\{ 
\left( \overline{\chi}^{cs}_x [\mu^{cs}_\kappa + i\gamma_5\tau_3a\mu_{cs}
-\tau_1a\mu_\delta ]\chi^{cs}_x \right)
- \half\sum_{\mu=\pm 1}^{\pm 4}
\left( \overline{\chi}^{cs}_{x+\hat{\mu}}U_{x\mu}[r+\gamma_\mu]\chi^{cs}_x \right)
\right\} \ ,
\ee
where we use the same notation as in eq.~(\ref{eq:ferm_action}), in particular
$\mu^{cs}_\kappa \equiv am^{cs}_0 + 4r = 1/2\kappa_{cs}$.
The new term with coefficient $\mu_\delta$ (the minus sign is conventional) 
is responsible for the splitting.
This construction preserves reality and, for $\mu_{cs}^2>\mu_\delta^2$,
positivity of the fermionic determinant.
The particle content of the theory is more transparent in a different basis
when $\chi_x\rightarrow\exp{\{i\frac{\pi}{4}\tau_2\}}\chi_x$,
$\bar\chi_x\rightarrow\bar\chi_x\exp{\{-i\frac{\pi}{4}\tau_2\}}$, where 
the mass term in the untwisted direction is diagonal and the twist is in the 
$\tau_1$ flavor direction:
\be\label{eq:plit_doub_phmt}
S^{cs}_{q, mass} = \sum_x  
 \overline{\chi}^{cs}_x [\mu^{cs}_\kappa +\tau_3a\mu_\delta+ i\gamma_5\tau_1a\mu_{cs}
 ]\chi^{cs}_x\ .
\ee
In analogy with the degenerate case the ``average'' quark mass of the split-doublet
is given by (we neglect here for simplicity renormalization factors)
$m_{cs}=\sqrt{(m^{cs}_0-{m^{cs}_{0\,cr}})^2+{\mu_{cs}}^2}$
and the individual masses by $m_{c,s}=m_{cs}\pm\mu_\delta$.
At full twist:
\be 
m_u=m_d=\mu_{ud}\ ,\;\; 
m_c=\mu_{cs}+\mu_\delta\ , \;\; 
m_d=\mu_{cs}-\mu_\delta \ .
\label{eq:qmasses_3}
\ee

\subsection{Tuning to full twist}

Tuning to full twist the $N_f=2+1+1$ theory looks apparently more complicated because 
of the presence of two tunable variables  $m^{ud}_0$ and $m^{cs}_0$ 
(or hopping parameters $\kappa_{ud}$ and $\kappa_{cs}$) 
with their respective critical values. This problem can be however circumvented 
in a strategy analogous to the $N_f=2$ case. 

As we have seen previously in this contribution for the 
$N_f=2$ theory, one possible definition of the critical 
quark mass ${m_0}_{cr}(g_0,\mu)$ is given by the vanishing of the corresponding 
PCAC quark mass $m^{PCAC}_\chi$. Due to chirality breaking the latter gets shifted:
\be
m^{PCAC}_\chi= m_0-a^{-1}f(g_0,am_0,a\mu)\ ,
\ee
with $f$ a dimensionless function.
On the basis of the symmetry of the action
under parity $\times$ $(\mu\rightarrow-\mu$) one can show that the additive renormalization
of the quark mass is {\em even} in $\mu$, and analyticity in turn implies
\be 
f(g_0,am_0,a\mu)= f(g_0,am_0)+O(\mu^2a^2)\ ,
\ee
where $f(g_0,am_0)$ is the shift for ordinary $N_f=2$ QCD without 
twisted mass term.
So the twisted mass term in the action only produces an $O(a)$ effect on the quark mass 
(with $g_0$ and $m_0$ held fixed):
\be\label{pcac_nf2}
m^{PCAC}_\chi=m_0-a^{-1}f(g_0,am_0)+O(a)\ .
\ee
The above argument can be easily generalized to the 
$N_f=2+1+1$ theory.
Here one has to make a distinction between the two sectors:
\bea
m^{PCAC}_{\chi(ud)} &=& m^{ud}_0-a^{-1}f_{ud}(g_0,am^{ud}_0,am^{cs}_0,a\mu_{ud},a\mu_{cs},a\mu_\delta) \ ,\\
m^{PCAC}_{\chi(cs)} &=& m^{cs}_0-a^{-1}f_{cs}(g_0,am^{cs}_0,am^{ud}_0,a\mu_{cs},a\mu_{ud},a\mu_\delta) \ .
\eea
The functions $f_{ud}$ and $f_{cs}$ are in this case even in $\mu_{ud}$, $\mu_{cs}$  and 
$\mu_\delta$\footnote{An additional symmetry in the $cs$ sector is needed for the argument, namely
$\chi_x\rightarrow\exp{\{i\frac{\pi}{2}\tau_1\}}\chi_x$,
$\bar\chi_x\rightarrow\chi_x \exp{\{-i\frac{\pi}{2}\tau_1\}}$ composed with 
$\mu_\delta\rightarrow-\mu_\delta$.}: similarly to $N_f=2$, the associated terms in the action only 
affect the additive renormalization of the quark mass by $O(a)$ terms. So we write:
\bea\label{eq:pcac_nf2p1p1_1}
m^{PCAC}_{\chi(ud)}&=&m^{ud}_0-a^{-1}f(g_0,am^{ud}_0,am^{cs}_0)+O(a) \ ,\\
m^{PCAC}_{\chi(cs)}&=&m^{cs}_0-a^{-1}f(g_0,am^{cs}_0,am^{ud}_0)+O(a)\ ,
\label{eq:pcac_nf2p1p1_2}
\eea
where on the r.h.s. we have now the mass-shifts for the theory without twist and mass-splitting
($N_f=2+2$ QCD): here the distinction between the two sectors is immaterial.
From eqs.~(\ref{eq:pcac_nf2p1p1_1}), (\ref{eq:pcac_nf2p1p1_2}) it follows immediately
\bea
 m^{ud}_0=m^{cs}_0=m_0 \quad \Rightarrow \quad m^{PCAC}_{\chi(cs)}=m^{PCAC}_{\chi(ud)}+O(a)\ .
\eea
The above result suggests to tune $m_0^{ud}$ to its critical value where $m^{PCAC}_{\chi(ud)}=0$
keeping  $m_0^{cs}=m_0^{ud}$ (or $\kappa_{cs}=\kappa_{ud}$): in this situation $m^{PCAC}_{\chi(cs)}=O(a)$. 
Observe that since the average quark mass in the $cs$ sector is typically large, the $O(a)$ error
is expected not to affect the full twist improvement in the sense of~\cite{Frezzotti:2003ni},
while it is critical to have good tuning in the light quark sector.

\subsection{Numerical simulations}

Our strategy is to simulate the $N_f=2+1+1$ theory at full twist. 
As previously explained for a given value
of $\beta$ this can be obtained by fixing $\kappa_{cs}=\kappa_{ud}=\kappa$ and tuning 
the PCAC quark mass in the light quark sector to zero (or equivalently by 
tuning the twist angle to $\pi/2$, 
see sec.~\ref{sec:omega_ren}). At the critical value $\kappa=\kappa_{cr}$,
$m^{PCAC}_{\chi(ud)}=0$ by definition and $m^{PCAC}_{\chi(cs)}=O(a)$, where the $O(a)$ 
error in the $cs$ sector is not expected to spoil $O(a)$ improvement. 

For the gauge action we choose the tree-level Symanzik improved action, 
which shows for $N_f=2$ an acceptable
behavior with respect to the phase transition at small quark masses and 
in addition presents several more advantages in comparison with the other 
possible candidate DBW2, see also section~5.

At present, we are 
simulating on two lattice-sizes: $12^3\cdot 24$ and $16^3\cdot 32$,
our preliminary objective being the reproduction of the physical situation
of previous simulation 
points~\cite{Farchioni:2004fs} for the $N_f=2$ theory. This means a lattice extension
$L= 2$~fm and $r_0/a=3$ and 4 
(corresponding to lattice spacings $a\simeq 0.17$~fm and $0.12$~fm, 
respectively), lightest pion mass $m_{\pi}\simeq 400$~MeV ($m_{ud}\simeq m_s/3$)
and $300$~MeV ($m_{ud}\simeq m_s/5$). 

The  $s$ and $c$ quark masses need to be tuned to their physical values. 
To this end we measure the kaon and $D$
meson masses and monitor the dimensionless quantities $r_0 m_K$ and $r_0 m_D$
which in nature assume the values $r_0 m_K\simeq 1.25$ and  $r_0 m_D\simeq 4.7$. 
Due to the $\tau_1$ term in the action, see eq.~(\ref{eq:plit_doub_phmt}), kaons and $D$ mesons
get mixed in the twisted theory. We solve the problem by diagonalizing
the two-by-two matrix of the $K$-$D$ correlators.

The analysis of the spectrum in the $cs$ mesonic sector ($D_s$) is 
even more awkward, again due to flavor mixing. The $\tau_1$ term in the action introduces indeed 
off-diagonal matrix elements in the quark propagator 
which in turn produce disconnected diagrams for the $D_s$ 
hadronic correlators; denoting with $\Delta^{ff^\prime}_{xy}$ the quark propagator, one has
$
C_{D_s (disc)}(x_0,y_0)=\sum_{\vec{x}} \left\langle {\rm Tr} \{ \Delta^{cs}_{xx}\gamma_5\}
                                       {\rm Tr} \{ \Delta^{sc}_{yy}\gamma_5\}\right\rangle \ .
$
This disconnected term is produced by $O(a)$ flavor symmetry breaking of the twist-term, 
similarly to that for the neutral pion in the light quark sector, cf. the contribution~\cite{pizero}
and ref.~\cite{Jansen:2005cg} 
for a discussion of this subject. 
(The present case is however a bit different, since the disconnected term originates 
from direct mixing between $c$ and $s$ quarks, while in the neutral pion case 
it is due to a mismatch
in the sign of the twisted mass for $u$ and $d$ quarks.) In this preliminary study we
measure the $D_s$ meson mass without taking into account the disconnected contribution.
\begin{table}
\centering
\begin{center}
\parbox{0.8\linewidth}{\caption{\label{tab:sim_par}
Simulation parameters of the performed simulations. See the text for the explanations 
of the symbols.
}}
\end{center}
\begin{center}
\begin{tabular}{c*{8}{c}}
 \hline
 \hline
 run & $\beta$ & $\kappa$ & 
 $a\mu_{ud}$ & $a\mu_{cs}$ & $a\mu_{\delta}$ & $N_{conf}$ \\
 \hline
 \hline
  ($a$) & 3.30 & 0.1720 & 0.01 & 0.325&0.275 
       & 550  \\
  ($b$) & 3.30 & 0.1725 & 0.01 & 0.325&0.275
       & 920  \\
  ($c$) & 3.30 & 0.1730 & 0.01 & 0.325&0.275 
       & 900  \\
  ($d$) & 3.30 & 0.1735 & 0.01 & 0.325&0.275
       & 730  \\
  ($e$) & 3.30 & 0.1740 & 0.01 & 0.325&0.275
       & 590  \\
 \hline
 \hline
\end{tabular}
\end{center}
\end{table}
\begin{table}
\centering
\begin{center}
\parbox{0.8\linewidth}{\caption{\label{tab:meas}
Measured quantities: the Sommer scale $r_0$; $\rho$ meson mass $m_\rho$, pion mass
$m_{\pi}$ and PCAC quark mass $am_{\chi(ud)}^{PCAC}$ (light quark sector); 
kaon, $D$ and $D_s$ meson masses. For the $D_s$ meson only connected diagrams 
have been taken into account.
}}
\end{center}
\begin{center}
\renewcommand{\arraystretch}{1.2}
\begin{tabular}{*{8}{l}}
  \hline 
  \hline 
  run & \,\,\,\,\,\,\, $r_0/a$ & \,\,\,\, $am_{\pi}$ & 
  \,\,\,\, $am_{\rho}$ & \,\,\,\, $am_{K}$ &
  \,\,\,\, $am_{D}$ & \,\,\,\, $am_{Ds}$ & 
  \,\, $am_{\chi(ud)}^{PCAC}$  \\  
  \hline
  \hline
  ($a$) & 2.23(7)(10) & 0.7102(38) & 0.8884(41) & 0.9551(21) &
   1.456(10) & 1.6456(49) & 0.08795(57)  \\
  ($b$) & 2.40(5)(5) & 0.6069(61) & 0.8075(38) & 0.9078(41) &
   1.420(11) & 1.6385(47) & 0.06399(73) \\
  ($c$) & 3.05(3)(8) & 0.3463(49) & 0.633(13) & 0.7940(16) &
   1.247(19) & 1.5684(59) & 0.01589(82) \\
  ($d$) & 4.11(6)(10) & 0.6012(88) & 0.981(14) & 0.9074(30) &
   1.119(18) & 1.4074(58) & -0.0711(21) \\
  ($e$) & 4.45(9)(20) &0.808(33) & 1.1429(47) &1.0004(17) &
   1.311(11) & 1.4086(54) &  -0.1208(46) \\
  \hline
  \hline
\end{tabular}
\end{center}
\end{table}

The simulation of the $N_f=2+1+1$ theory is performed by a polynomial hybrid Monte Carlo 
algorithm (PHMC)~\cite{Frezzotti:1997ym}. The structure of the algorithm goes along the
lines indicated in~\cite{Montvay:2005tj}. In table~\ref{tab:sim_par} we report the parameters
of the first simulations performed. The parameters were chosen in such a way 
that the bare quark masses follow at full twist the proportions $m_u:m_d:m_s:m_c=1:1:5:60$, 
see~(\ref{eq:qmasses_3})  
(of course renormalizations can alter these proportions).
The point at $\kappa=0.1730$ is the one nearest to full twist where the pion mass is minimal:
here $r_0/a\simeq 3$, $m_\pi\simeq 400$~MeV and the $D$ meson mass is near to 
the physical value, but the kaon is heavier than in nature: $r_0m_K\simeq 2.4$. 
So the next step will be to decrease the $s$ quark mass. 
Simulations on a $16^3\cdot 32$ lattice have been also started.

One interesting subject of investigation in perspective is of course the first 
order phase transition at small quark masses in this new $N_f=2+1+1$ setup, and
in particular how it compares with previous studies with different 
gauge actions and/or quark flavor content ($N_f=2$ and Wilson plaquette, 
DBW2 or tlSym gauge actions). 
The question arises, how the dynamics of the $s$ quark can influence the 
phase transition and, more in general, the low-energy features of QCD. 
For the moment we can only state that no signs of metastabilities were found in 
the present runs, which could however be just too far from the metastable region. 

\section{Conclusions}

We have investigated the phase structure of lattice QCD by probing different gauge 
actions, adding to the standard Wilson plaquette action
a term proportional to the rectangular plaquette, see 
eq.~(\ref{eq:gauge_action}). 
By varying the coupling $c_1$ which multiplies the 
rectangular plaquette term, one can
interpolate between various actions and this allows to understand in
more detail the properties of the phase structure, in particular how the
strength of the transition depends on the additional term and how
this influences the approach to the continuum limit.
For the fermion action we always used Wilson twisted mass fermions at 
various values of the twisted mass parameter $\mu$.

\begin{figure}[htb]
\vspace{-0.0cm}
\begin{center}
\includegraphics[width=0.6\linewidth]{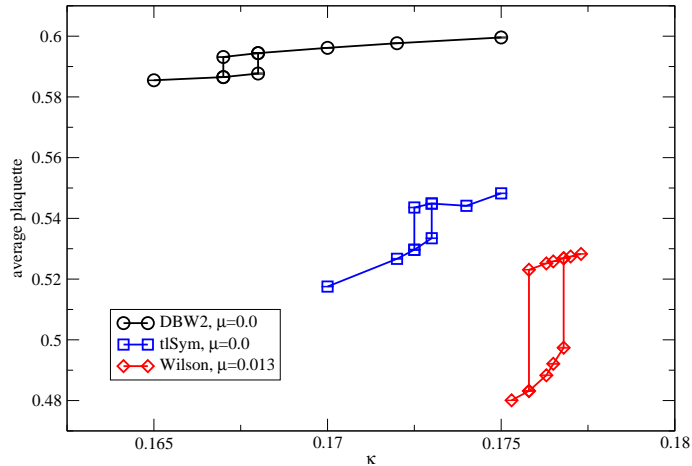}
\end{center}
\vspace{-0.7cm}
\caption{Hysteresis of the average plaquette value as $\kappa$ is moved across
  the critical point, for Wilson, tlSym and DBW2 gauge action at $a\sim 0.17$~fm.
\label{fig:comp_plaq_wilson_dbw2_tlsym}}
\end{figure}

That even a small value of $c_1$ can already
have a large impact on the phase structure is illustrated in 
figure \ref{fig:comp_plaq_wilson_dbw2_tlsym}
where we show the average plaquette value as a function of the hopping
parameter $\kappa$ for three different actions, i.e.~values of $c_1$, namely
$c_1=0$ (Wilson), $c_1=-1/12$ (tlSym) and $c_1=-1.4088$ (DBW2).
As one moves $\kappa$ from the negative or positive side across the critical
point, where the PCAC quark mass vanishes, a hysteresis in the average
plaquette value develops of which the size and the width is an indicator of
the strength of the phase transition. We observe that both the width and the
size of the gap in the plaquette value decreases considerably as we switch on
$c_1$ to $c_1=-1/12$ (tlSym action).  Decreasing $c_1$ further down to
$c_1=-1.4088$ (DBW2 action) still seems to reduce the size of the gap, but the
effect is surprisingly small despite the large change in $c_1$. Note that the
results in figure \ref{fig:comp_plaq_wilson_dbw2_tlsym} are for a lattice
spacing $a\sim 0.17$~fm that is roughly consistent for all three actions.

However, the results for the Wilson plaquette gauge action are at non-zero twisted mass
$\mu=0.013$, while they are at zero twisted mass for the tlSym and the DBW2
action. Since the strength of the phase transition is expected to be reduced
as one switches to a non-zero twisted mass, a true comparison at $\mu=0$ would
disfavor the Wilson plaquette gauge action even more. Note also that the phase
transition is a generic feature of dynamical Wilson fermion simulations
independent of whether a twisted mass is switched on, i.e.~the transition also
occurs for standard Wilson fermion simulations.
Another feature of the first order phase transition is that 
its strength weakens rapidly when the lattice spacing is made finer.
This is illustrated in fig.~\ref{fig:plaq} for the case of the 
Wilson plaquette gauge action. 

A satisfactory setup for dynamical simulations with, say, $N_f=2$ flavors of
quarks  would be to reach pion masses of about $250-300$~MeV and a box size
of $L>2$~fm. At the same time, one should stay at full twist to realize 
O(a)-improvement.
From our preliminary results we find that for the tlSym action this can be
achieved with a reasonable computer time at $\beta=3.9$ on $L/a=20$ lattices. 
For smaller values of $\beta$ we find that at pion masses of about 
$400$~MeV large fluctuations appear, although no clear signs of 
metastabilities are visible.
Although for the DBW2 action the situation might be somewhat better, the 
advantages of the tlSym action such as good convergence of perturbation 
theory and small scaling violations as found quenched, lead us to decide on 
the tlSym gauge action as the action of choice. 

For our present results from simulations using the Wilson plaquette and the DBW2
gauge actions we found that $\chi$PT describes the data 
rather well and allows to extract a number of low-energy constants of 
the effective chiral Lagrangian. We have also seen that scaling violations
of both actions are not very large when we compared physical results 
at different values of the lattice spacing. 

Very good news is that simulations with $N_f=2+1+1$ flavors of dynamical quarks
are perfectly feasible with the twisted mass formulation of lattice QCD. 
Our first and preliminary results are very promising. The (PHMC) algorithm 
works nicely and, even more important, it seems that the tuning in this 
physically more realistic setup is quite manageable. Taken this fact together 
with the new developments of fast  
versions of the HMC algorithm to simulate 
dynamical quarks, 
it seems natural to start now realistic simulations with twisted mass
fermions.

{\large\bf Acknowledgments}

We thank H.~Perlt for providing us with the
perturbative estimates of the renormalization constants of the quark
bilinears for the DBW2 gauge action.
We thank R.~Frezzotti, G.~C.~Rossi and S.~Sharpe for many useful
discussions. The computer centers at NIC/DESY Zeuthen, NIC at
Forschungszentrum J{\"u}lich and HLRN provided the necessary technical
help and computer resources. We are indebted to R.~Hoffmann and
J.~Rolf for leaving us a MatLab program to check our fits.
This work was supported by the DFG Sonderforschungsbereich/Transregio
SFB/TR9-03.

\bibliographystyle{JHEP-2}
\bibliography{proc}
\end{document}